\documentclass[letter]{aa}

\usepackage{txfonts}
\usepackage{CJK}
\usepackage{caption}
\usepackage{subcaption}
\usepackage{dcolumn}
\usepackage{lscape}
\usepackage{amsmath}
\usepackage{booktabs}
\usepackage{color, colortbl}
\definecolor{ratgray}{gray}{0.9}
\definecolor{dblue}{rgb}{0.0, 0.0, 0.65}
\usepackage[colorlinks = true,linkcolor = blue,citecolor = blue]{hyperref}

\begin{document} 
\begin{CJK}{UTF8}{gbsn}
   \title{An Extreme-AO Search for Giant Planets around a White Dwarf\thanks{Based on observations made with European Southern Observatory (ESO) telescopes at the La Silla Paranal Observatory under program 60.A-9373(A).}
   }

   \subtitle{VLT/SPHERE performance on a faint target GD 50}

   \author{
      S. Xu (许\CJKfamily{bsmi}偲\CJKfamily{gbsn}艺)\inst{1}
          \and
      S. Ertel\inst{2}
          \and
      Z. Wahhaj\inst{2,3}
          \and
      J. Milli\inst{2}
          \and
      P. Scicluna\inst{4}
          \and
      G. H.-M. Bertrang\inst{4}
           }

   \institute{
      European Southern Observatory, Karl-Schwarzschild-Stra{\ss}e 2, 85748
Garching, Germany\\
      \email{sxu@eso.org}
          \and
      European Southern Observatory, Alonso de Cordova 3107, Vitacura, Casilla
19001, Santiago 19, Chile
	\and
Aix Marseille Universite, CNRS, LAM (Laboratoire d'Astrophysique de Marseille) UMR 7326, 13388, Marseille, France
          \and
      Kiel University, Institute of Theoretical Physics and Astrophysics,
Leibnizstr. 15, 24118 Kiel, Germany
              }

   \date{Received 24 March, 2015; accepted 8 June, 2015}

 
  \abstract
   {  Little is known about the planetary systems around single white dwarfs although there is strong evidence that they do exist.  }
   {We performed a pilot study with the extreme-AO system on the Spectro-Polarimetric High-contrast Exoplanet REsearch (SPHERE) on the Very Large Telescopes (VLT) to look for giant planets around a young white dwarf, GD 50.
   }
   {We were awarded science verification time on the new ESO instrument SPHERE. Observations were made with the InfraRed Dual-band Imager and Spectrograph in classical imaging mode in H band.
   }
   {
  Despite the faintness of the target   (14.2 mag in R band), the AO loop was closed and a strehl of 37\% was reached in H band. No objects were detected around GD 50. We achieved a 5-sigma contrast of 6.2, 8.0 and 8.25 mags at 0{\farcs}2, 0{\farcs}4 and 0{\farcs}6 and beyond, respectively. We exclude any substellar objects more massive than 4.0 M$_\textrm{J}$ at 6.2 AU, 2.9 M$_\textrm{J}$ at 12.4 AU and 2.8 M$_\textrm{J}$ at 18.6 AU and beyond. This rivals the previous upper limit set by Spitzer. We further show that SPHERE is the most promising instrument available to search for close-in substellar objects around nearby white dwarfs.}
   {}


   \maketitle

\section{Introduction}

Radial velocity and transit surveys demonstrate that extrasolar planets are prevalent around main-sequence stars in the Milky Way \citep{Cumming2008}. However, little is known about the evolution of planetary systems beyond the main sequence life time of their host stars. Planets have been discovered around (sub)giant stars \citep{Johnson2011, Gettel2012} as well as horizontal branch stars \citep{silvotti2007, setiawan2010, charpinet2011}. There is only one planetary mass object detected with direct imaging around a white dwarf (\mbox{WD 0806-661}; \citealt{Luhman2011}) although theoretical calculations show that many more planets can survive the red giant stage \citep{Burleigh2002, Jura2008, Mustill2012, NordhausSpiegel2013}.

From an observational perspective, the strongest evidence for planets around single white dwarfs comes from the study of heavy-element-enriched white dwarfs. About 25-50\% of white dwarfs have heavy elements in their atmospheres, in addition to the primordial hydrogen or helium \citep{Zuckerman2010, Koester2014}. There must be a constant supply of material onto the white dwarf to balance the rapid settling of heavy elements out of the photosphere \citep{Koester2009a}. In addition, some of these heavy-element-enriched white dwarfs also display infrared excess from a circumstellar dust disk \citep{Jura2007b, XuJura2012}, or more rarely a gaseous disk \citep{Gaensicke2006}. The standard model is that these were formed by asteroids that were perturbed by the surviving planet(s) in the system onto orbits that crossed the tidal radius of the white dwarf, were disrupted, and formed a dust disk; eventually, these materials were accreted onto the white dwarf, enriching its atmosphere \citep{Jura2003, JuraYoung2014}. This asteroid accretion model is further supported by comparing the abundances of the accreting material with solar system objects -- they all resemble terrestrial material rather than the cosmic abundance \citep{Klein2010, Jura2012, Gaensicke2012, Xu2014}. 

Dynamical simulations show that there are different pathways that can lead to the asteroid accretion model. The configuration varies between having a comet belt with two planets -- an inner planet and a Neptune-sized planet \citep{Bonsor2011}, an asteroid belt with a Jupiter-sized planet \citep{Debes2012a} and an asteroid belt with an eccentric sub-Jupitzer-sized planet \citep{FrewenHansen2014}. At least one planet must be present to explain all the observed features of the system.

However, classical planet-hunting methods do not work well for white dwarfs. The red giant star has cleared planets within a few AUs, making the transit probability very low. White dwarfs have far too few spectral lines to make accurate radial velocity measurements.  Here, we summarize a few previous searches for planets arounds single white dwarfs. 

\begin{itemize}

\item {\it Direct imaging.} It has long been speculated that direct imaging should be the most promising method to detect planets around white dwarfs due to: (i) the better contrast between the white dwarf and the potential planet compared to main sequence stars; (ii) the orbital expansion of the planet during the red giant stage \citep{Burleigh2002, GouldKilic2008}. The only planetary mass object around a single white dwarf was discovered via direct imaging with Spitzer; \mbox{WD 0806-661B} has a mass of 7 M$_\textrm{J}$ at an orbital separation of 2500 AU \citep{Luhman2011}. The most systematic search is the DODO (Degenerate Objects around Degenerate Objects) survey, which observed 29 white dwarfs with Gemini/NIRI and VLT/ISAAC; their average upper limit is $\sim$ 8$M_\textrm{J}$ beyond 35 AU \citep{Hogan2011}. There are also a few other direct imaging studies using HST/NICMOS \citep{Debes2005a, ZinneckerKistionas2008}.

\item {\it Infrared excess.} Because a white dwarf's luminosity peaks in the ultraviolet or optical, a cool companion can be easily detected as an infrared excess. Those searches started in the late 80s and so far, only brown dwarf companions have been discovered \citep{ZuckermanBecklin1987, Maxted2006, Farihi2008a, Kilic2009c}.

\item  {\it Planetary transit.} \citet{Faedi2011} studied 194 white dwarfs in the WASP survey and put an upper limit of 10\% for Jupiter-sized brown dwarfs and giant planets with orbital periods less than 0.1-0.2 d. \citet{Fulton2014} searched for eclipses around $\sim$ 1700 white dwarfs in the Pan-STARR1 field and found zero candidates, giving an upper limit of 0.5\% for Jupiter-sized-planet orbiting just outside of the Roche limit. 

\item {\it White dwarf pulsation.} Some white dwarfs have extremely stable pulsation modes and the presence of a planet can cause a periodic change in the arrival time of the pulsation. \citet{Mullally2008} led a study on 15 such stable pulsators and identified one candidate -- a 2$M_\textrm{J}$ mass planet in a \mbox{4.5 yr} orbit around GD 66. The planet hypothesis was dismissed later due to some other peculiar behavior in the pulsation of \mbox{GD 66} \citep{Mullally2009, Farihi2012, Hermes2013}. The upper limit from pulsation studies is typically a few Jupiter masses within 5 AU of the white dwarf \citep{Mullally2008}.

\end{itemize}

In this paper, we present a pilot search for giant planets around a young white dwarf GD 50. The data were taken in a science verification (sv) program with the new extreme-AO instrument SPHERE \citep{Beuzit2008} at ESO's VLT. SPHERE includes the SAXO extreme adaptive optics system \citep{Fusco2014, Petit2014}, with a deformable mirror of 41 $\times$ 41 actuators, and 4 control loops (tip-tilt, high-orders, near-infrared differential tip-tilt and pupil stabilization). 

\section{Observation and Data Reduction}

\subsection{Target Selection}

GD 50 (WD 0346-011) is a young nearby ultramassive hydrogen atmosphere white dwarf (m$_\textrm{V}$ = 14.0, m$_\textrm{R}$=14.2, m$_\textrm{H}$ = 14.9). Its formation scenario has puzzled many astronomers. \citet{Dobbie2006b} performed a complete astrometric and spectroscopic analysis and concluded that it has evolved as a single star and is an escaping white dwarf from the Pleiades open cluster. By fitting the photometric and spectroscopic data, they derived the effective temperature, surface gravity and distance of GD 50 to be T = 41550 K, log g = 9.15 and d = 31 pc. This corresponds to a white dwarf with a mass of 1.264 M$_\odot$ and a total age of 125 Myr, consistent with the age of Pleiades. According to the initial-final mass relation in \citet{Williams2009}, it had an initial mass of \mbox{ 7 M$_{\odot}$,} making its progenitor a B-type star.

GD 50 was chosen as our target for the SPHERE sv program due to its youth and close proximity. At 125 Myr, potential planets can still be quite hot and thus detectable. In addition, \mbox{GD 50} has been observed with Spitzer/IRAC and its 3 $\sigma$ upper limit of \mbox{$\sim$ 4 M$_\textrm{J}$}\footnote{\citet{Farihi2008a} reported a 3 $\sigma$ upper limit of  3 M$_\textrm{J}$ assuming a total age of 100 Myr for GD 50. To make the results directly comparable with our SPHERE observation, we adopted the revised age of 125 Myr and the updated BT-Settl model grid \citep{Allard2014}. Our updated Spitzer upper limit is 4.2 M$_\textrm{J}$ for GD 50. } is among the best constraints from Spitzer searches \citep{Farihi2008a}. A side goal of the project is to compare the performance of SPHERE with other instruments for detecting substellar objects around white dwarfs.

\subsection{Observation}

GD 50 was first observed on December 5th, 2014. Due to poor observing conditions, the data were not used for the analysis. The observation was repeated on February 1st, 2015 with seeing around 1{\farcs}0, a moon distance of $46^\circ$ (89\% illumination) and a coherence time of $\sim$ 5 ms. The InfraRed Dual-band Imager and Spectrograph (IRDIS) \citep{Dohlen2008} was used in classical imaging (CI) mode in broadband H. The observations were made in pupil-stabilized mode, ideal for PSF subtraction, as the pupil and PSF stay fixed while the field rotates with respect to the detector. The integration time was 64 sec per exposure and the observing block consisted of a $4\times4$ dither pattern with 3 exposures at each position, giving a total on-source time of 3072 sec.

\subsection{Data Reduction \label{DataReduction}} 

\begin{figure}[h]
    \includegraphics[angle=0,width=0.8\linewidth]{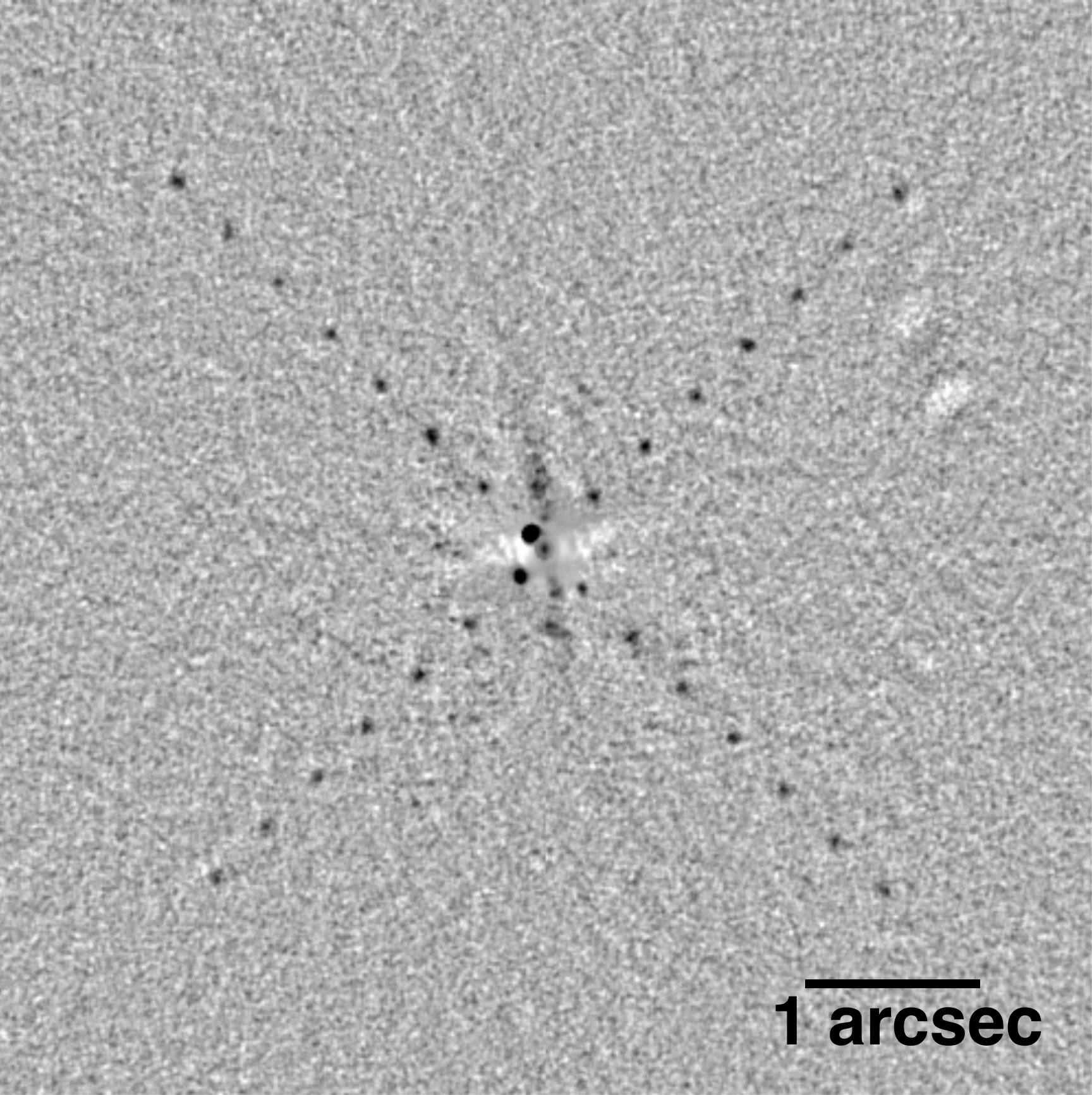}
 \caption{Signal-to-noise map (i.e., reduced image divided by a noise map derived from it, see Section \ref{DataReduction}). The map has been smoothed using a Gaussian of FWHM 1.5 pixels. No astronomical point-like features were detected. In addition, we recovered artificial sources inserted into the raw data at the 5$\sigma$ contrast level at separations ranging from 100 mas to 3 arcsec with increments of 100 mas (X shaped pattern of sources). The field of view displayed here is 6{\farcs}25 by 6{\farcs}25. Counts are displayed in linear scale and inverted (dark corresponds to high counts). North is up. The central white dwarf  is not visible in this map because of image processing.
 }
      \label{Image}
\end{figure}

The raw images were first corrected for flatfield errors and bad pixels.  The white dwarf was unsaturated in all the images. Since no coronagraph was used, we were able to align the star on all images with subpixel accuracy. Due to the faintness of the star, the stellar halo is weak and there are hardly any speckles. Thus there is no need for PSF-subtraction. All images were derotated to align North up and median combined. Following \citet{Wahhaj2013}, a filtering process was applied to the median-combined image to isolate spatial features of $\sim$ \mbox{50 mas} in size, the diffraction limit in H band. This filter removes azimuthally extended features, e.g. the stellar halo, and improves the sensitivity to point sources for separations less than 0{\farcs}3. The signal-to-noise map is shown in Fig. \ref{Image}. 

\begin{figure}[h]
  \includegraphics[angle=0,width=\linewidth]{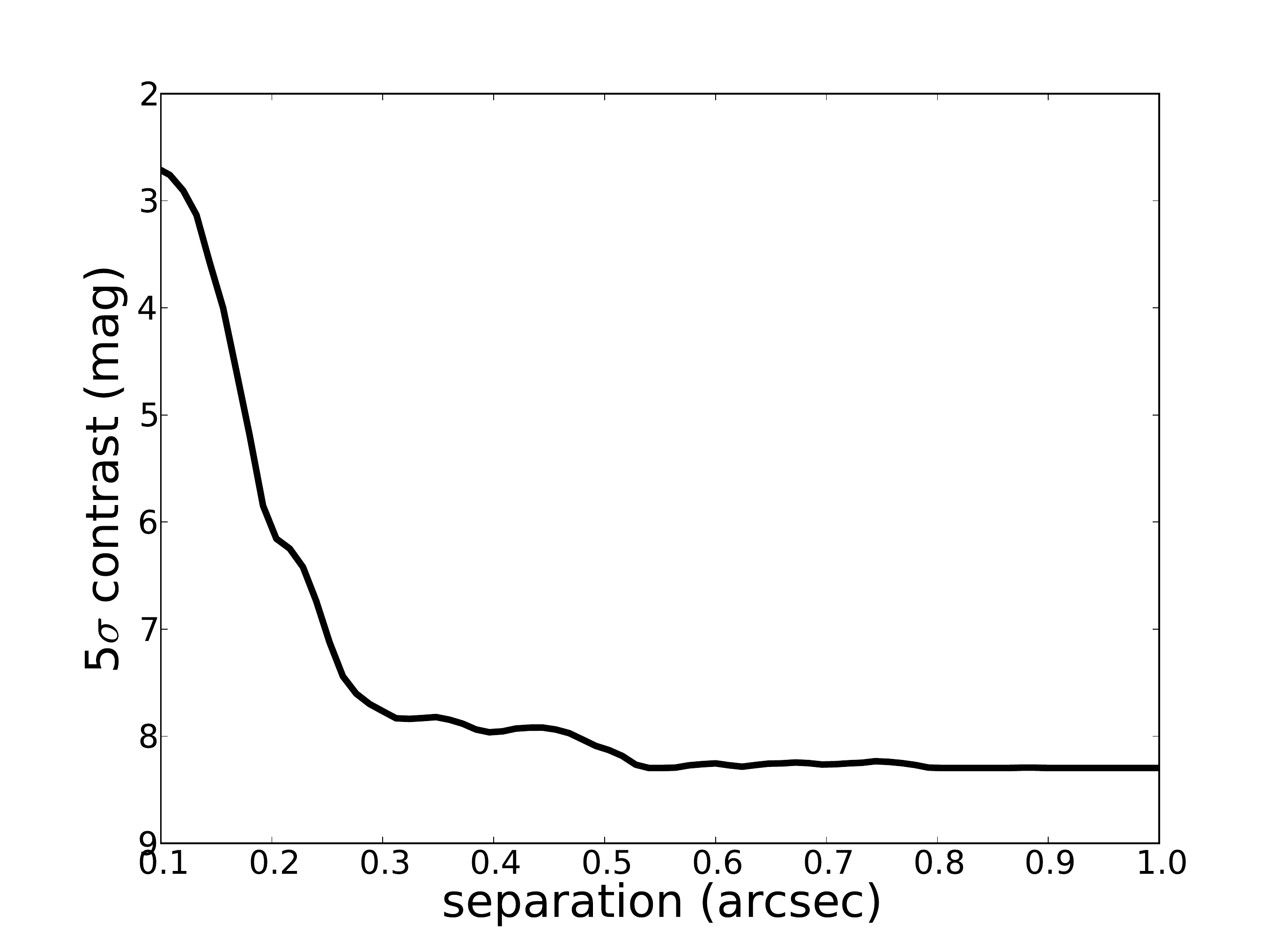}
 \caption{5$\sigma$ contrast curve for GD 50. A 5$\sigma$ contrast of 6.2, 7.8, 8.0, 8.1 and 8.25 mags has been reached at 200 mas, 300 mas, 400 mas, 500 mas and 600 mas, respectively.
 }
 \label{Contrast}
\end{figure}

Then we calculated the 5$\sigma$ contrast curve, or the sensitivity limit relative to the primary. The peak intensity (star peak) at the stellar centroid was estimated by cubic interpolation. The pixel-to-pixel RMS was then measured in annuli of 5 pixels radii centered on the star for all separations (radii) of interest. The 5$\sigma$ contrast was computed as \mbox{(star peak)/(RMS $\times$ 5.0)}, where RMS is a function of separation from the star. No sources except the white dwarf were detected and the contrast curve is shown in Fig. \ref{Contrast}. 

To check the robustness of our sensitivity limit,  we cut out a circular aperture of radius 4 pixels around the star from each basic reduced image and re-inserted it at the 5$\sigma$ contrast limit at several separations and position angles in the individual images. All the images were derotated and median-combined again. As shown in Fig. \ref{Image} , all the inserted companions are recovered in the median combined image. We have achieved 5$\sigma$ contrasts of 6.2, 7.8, 8, 8.1 and 8.25 mags at 0{\farcs}2, 0{\farcs}3, 0{\farcs}4, 0{\farcs}5 and 0{\farcs}6, respectively. In the background limited region, the $5\sigma$ limit corresponds to an apparent magnitude in H band of m$_\textrm{H}$ = 23.15 (absolute magnitude M$_\textrm{H}$ =20.7).

\section{Discussion}
\label{sect_discussion}

\subsection{Upper limits on planetary mass companions of GD\,50}
\label{sect_masses}

\begin{figure}[h]
 \centering
 \includegraphics[angle=0,width=\linewidth]{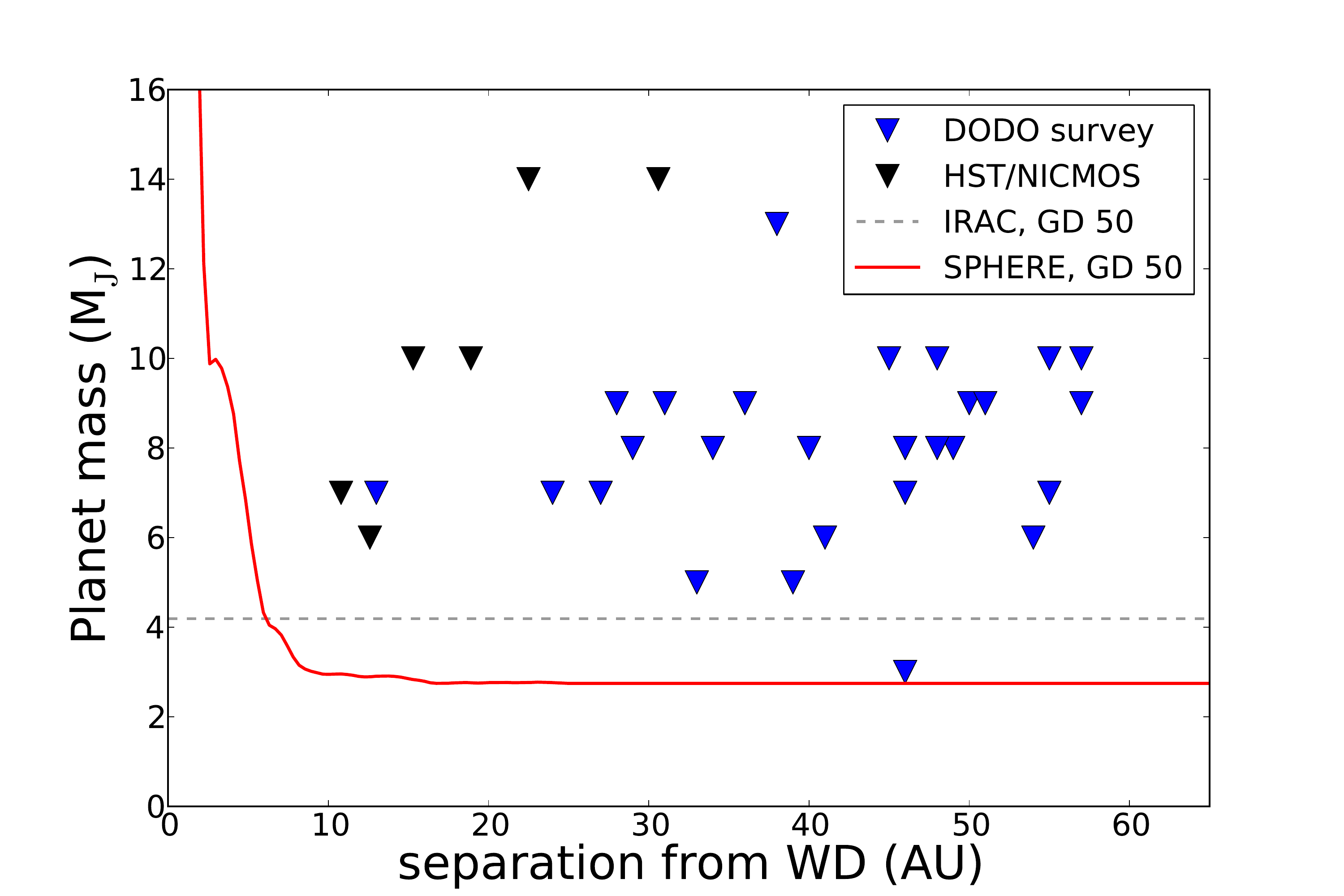}
 \caption{ Upper limits from previous imaging surveys (DODO survey from \citealt{Hogan2011}, HST/NICMOS studies from \citealt{Debes2005a}) for substellar objects around white dwarfs. The only positive detection so far is WD 0806-661B, but it is not shown because the separation is 2500 AU \citep{Luhman2011}. We also include the upper limit on GD 50 from Spitzer/IRAC studies \citep{Farihi2008a}. GD 50 is the most distant white dwarf in the figure, yet our limit on GD 50 is still the most stringent.
 }
 \label{WD_Planet}
\end{figure}

To convert our detection limit to planetary mass, we used the BT-Settl models from \citet{Allard2014}. We took the projected separation as the orbital distance from the white dwarf and did not correct for the inclination effect. As shown in Fig. \ref{WD_Planet}, we put an upper limit to the planetary mass of 4.0 M$_\textrm{J}$, 2.9 M$_\textrm{J}$ and \mbox{2.8 M$_\textrm{J}$} at 6.2 AU, 12.4 AU and at 18.6 AU and beyond, respectively.

Our upper limit corresponds to a limit of 2.8 M$_\textrm{J}$ beyond \mbox{3.4 AU} during the main-sequence stage of the star. Depending on the metallicity, a 7M$_\odot$ star could reached a maximum radius about 6.5 AU during the post main-sequence evolution \citep{Veras2013}, so any planets within 6.5 AU would possibly have been destroyed. Therefore, it is unlikely that there are any giant planets more massive than 2.8 M$_\textrm{J}$ around \mbox{GD 50} now. For the surviving planets, their orbits will expand by a factor of about 7$/$1.264, which will be at least 36 AU from GD 50. However, tidal interaction between the planet and the extended envelope of the AGB star can counteract the orbital expansion \citep{Mustill2012}. Furthermore, if multiple planets were present, mutual interactions could scatter planets inward during the white dwarf stage \citep{VerasGaensicke2015}. As a result, some planets might end up having a much smaller semi-major axis. In addition, no heavy elements have been detected in the atmosphere of \mbox{GD 50}; though optical observations are not very sensitive to heavy elements at high stellar temperatures \citep{Koester2009b}. GD 50 has a temperature of 41550 K and the upper limits are not very constraining.

In Fig.~\ref{WD_Planet}, we compare the upper limit on GD 50 with all previous imaging searches for substellar objects around white dwarfs. Our upper limit on GD 50 is the most stringent, both in terms of mass and separation. For example, HST/NICMOS has an inner working angle of 0{\farcs}9 with the coronagraph \citep{Debes2005a}, which for GD 50 corresponds to 28 AU in Fig. \ref{WD_Planet}.

\subsection{Future observational prospects \label{future}}

\begin{figure*}[hptb!]
 \centering
 \includegraphics[angle=0,width=0.9\linewidth]{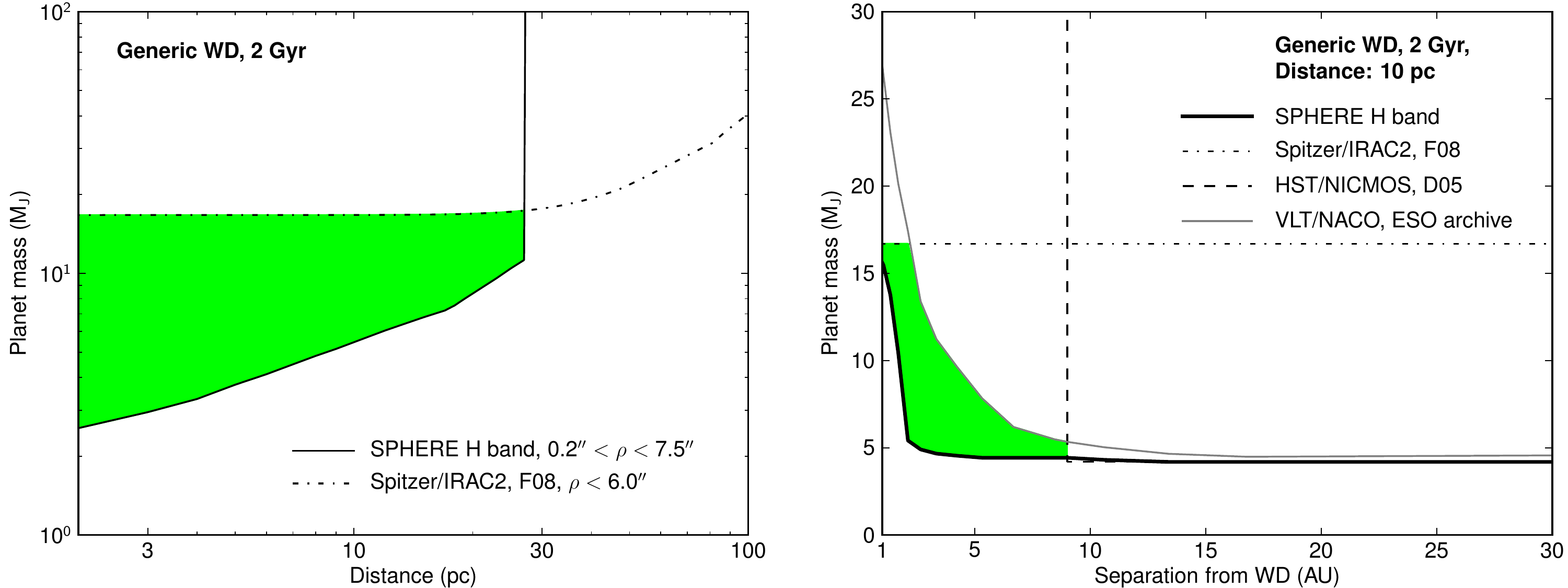}
 \caption{
 Illustration of the performance of SPHERE compared to other instruments for a typical nearby white dwarf (age 2\,Gyr, M$_\textrm{V}$ = M$_\textrm{H}$ = 12). The green regions represent the parameter space for which SPHERE enables significant improvement over the other instruments. \textit{Left}: Sensitivity vs. distance of the white dwarf from the Sun. The quantity $\rho$ represents the angular separation that the instrument is sensitive to. The performance of SPHERE is extrapolated from our observation of GD 50. Because GD 50 is at the faintness limit, the actual performance on a brighter target would be better. The vertical cut-off in the SPHERE curve near 30 pc is artificial and assumes that a star fainter than m$_\textrm{V}$=14.5 cannot be observed due to the limiting magnitude of the AO. For nearby bright white dwarfs, resolved imaging is more sensitive to low mass objects than photometric observations.
\textit{Right}: Contrast curves for the same WD at a distance of 10\,pc. The NACO curve is estimated from archival data (program ID: 079.D-0561(A), PI: Radiszcz) with a similar integration time as SPHERE (3000 sec). The HST/NICMOS limit was extrapolated from \citet{Debes2005} with a typical exposure time of 20 min. The key point is that SPHERE has the unique capability to detect high-mass planets at 2-10 AU from the white dwarf, which is not possible with any other instrument.
  }
 \label{estimate}
\end{figure*}

GD\,50 is not representative of nearby white dwarfs  -- it is particularly young. Now we investigate the performance of SPHERE compared to other instruments for a generic white dwarf within 20 pc, with an absolute magnitude M$_\textrm{V}$ = 12.0, M$_\textrm{H}$ = 12.0 and a total age of 2 Gyr \citep{Holberg2008b}. Since there are no BT-Settl models available for low-mass planets at 2 Gyr, AMES-COND models are used instead \citep{Allard2001}. As shown in the left panel of Fig.~\ref{estimate}, for nearby white dwarfs (within $\sim$ 30 pc), Spitzer photometric observation is limited by the calibration uncertainty of approximately 15\% ($3\sigma$); as a consequence, the Spitzer upper limit to planet mass remains the same regardless of the white dwarf distance. This is not the case for SPHERE because a potential planet would be spatially disentangled from the star outside an angular separation of $\sim$ 0{\farcs}2, about 2 AU at \mbox{10 pc}. Resolved imaging with Spitzer can be more sensitive to lower mass objects, but at a much larger orbital separation (e.g. WD 0806-661B in \citealt{Luhman2011}). Spitzer observations become background limited for more distant white dwarfs (beyond 30 pc) and all those white dwarfs are typically too faint for SPHERE. For nearby white dwarfs, resolved imaging with SPHERE is more sensitive to close-in low mass objects than photometric observations. There are $\sim$ 50 white dwarfs brighter than 14.5 mag in R band that are observable with SPHERE \citep{Holberg2008b}. They are prime targets for future searches for planetary systems with SPHERE.

In the right panel of Fig.~\ref{estimate}, we show the performance of SPHERE compared to other direct imaging instruments\footnote{The performance of GPI is not shown because it has a limiting magnitude of m$_\textrm{I}$ = 10 \citep{Macintosh2014}. There are very few white dwarfs that are bright enough to be observed with GPI. }. NICMOS and NACO have a much bigger inner working angle. Note that the difference between NACO, HST, and SPHERE in the background limited regime is minor. The real improvement of SPHERE over other instruments is in the region less than 1{\farcs}0. For a typical white dwarf within 20 pc, this corresponds to within 20 AU from the star. This region is particularly interesting, because it can constrain models of the survival of planetary systems during the AGB phase. 

Most white dwarfs have a progenitor mass between 2-5 M$_{\odot}$. Searching for planets around white dwarfs directly probes this intermediate-mass region, which is very difficult with radial velocity and transit methods \citep{Galland2005}. Recently, \citet{Barber2014} discovered a 1.04 M$_{\odot}$ white dwarf (progenitor mass 5.4 M$_\odot$) with a dust disk from tidally disrupted asteroid(s) -- the first confirmation that massive stars can have planetary systems.

\section{Summary}

We present a pilot study with the extreme-AO instrument SPHERE to search for giant planets around a young white dwarf, GD 50. We exclude any substellar objects above 4.0 M$_\textrm{J}$, 2.9 M$_\textrm{J}$ and 2.8 M$_\textrm{J}$ at 6.2 AU (0{\farcs}2), 12.4 AU (0{\farcs}4) and at 18.6 AU (0{\farcs}6) and beyond, respectively -- among the best constraints compared to all previous studies. We further show that SPHERE is the best instrument available to search for close-in substellar objects around nearby white dwarfs. 

\begin{acknowledgements}
The authors thank an anonymous referee for useful comments. We thank France Allard for discussing various giant planet models and Monika Petr-Gotzens for useful discussions of the manuscript. We thank Carl Melis and Jay Farihi for helpful suggestions. We thank the SPHERE science verification team for performing the observations.
\end{acknowledgements}

\bibliographystyle{aa}

\clearpage
\end{CJK}

\end{document}